\documentclass[%
10pt,twocolumn,letterpaper,aps,prl,superscriptaddress,showpacs,amsmath,
]{revtex4-1}

\usepackage{graphicx}

\newcommand\ket[1]{\lvert #1 \rangle}

\newcommand\moment[1]{\langle #1 \rangle}
\newcommand{\UT}{%
Department of Applied Physics, School of Engineering, \\
The University of Tokyo, 7-3-1 Hongo, Bunkyo-ku, Tokyo 113-8656, Japan
}
\newcommand{\UP}{%
Department of Optics, Palack\'y University,
17. listopadu 1192/12, 77146 Olomouc, Czech Republic
}
\newcommand{\SEIT}{%
School of Engineering and Information Technology, \\
University of New South Wales, Canberra, ACT 2600, Australia
}

\begin{document}

\title{%
Experimental realization of a dynamic squeezing gate
}

\author{Kazunori Miyata}
\affiliation{\UT}
\author{Hisashi Ogawa}
\affiliation{\UT}
\author{Petr Marek}
\affiliation{\UP}
\author{Radim Filip}
\affiliation{\UP}
\author{Hidehiro Yonezawa}
\affiliation{\SEIT}
\author{Jun-ichi Yoshikawa}
\affiliation{\UT}
\author{Akira Furusawa}
\affiliation{\UT}

\begin{abstract}
Squeezing is a nonlinear Gaussian operation that is the key component in construction of other nonlinear Gaussian gates. In our implementation of the squeezing gate, the amount and the orientation of the squeezing can be controlled by an external driving signal with 1 MHz operational bandwidth. This opens a brand new area of dynamic Gaussian processing. In particular, the gate can be immediately employed as the feed-forward needed for the deterministic implementation of the quantum cubic gate, which is a key piece of universal quantum information processing.
\end{abstract}

% 03.67.Lx Quantum computation architectures and implementations
% 42.50.Dv Quantum state engineering and measurements
% 42.50.Ex Optical implementations of quantum information
%          processing and transfer
% 42.65.-k Nonlinear optics

\maketitle

%%%%%%%%%%%%%%%%%%%%%%%%%%%%%%%%%%%%%%%%%%%%%%%%%%%%%%%%%%%%%%%%%%%%%%%%%%%%%%%%

Quantum information processing with continuous variable systems (CV) has many tools. They could be divided into two broad categories --- Gaussian and non-Gaussian. The Gaussian tools comprise of Gaussian quantum states that can be represented by a Gaussian Wigner function, of Gaussian measurements that project on Gaussian states, and of Gaussian operations that transform Gaussian states into different Gaussian states \cite{Hybrid}. The non-Gaussian tools category then includes everything else. The non-Gaussian category is much broader and much more powerful. There are many quantum information protocols that cannot be implemented with Gaussian tools alone, quantum computation \cite{QcOverCv,CvCluster}, entanglement distillation \cite{Qdist1,Qdist2,Qdist3}, and error correction \cite{Gkp} are just the three most prominent examples.

As a consequence, there is an understandable thirst for all matters non-Gaussian. In quantum optics, which is the experimental platform of choice when it comes to tests of CV paradigms \cite{QiWithCv}, the non-Gaussian features need to come from interactions with discrete variable physical systems \cite{ProbingCavityWithAtom, CrossPhaseModWithAtom, SinglePhotonLevelNonlinearityWithAtom, PhaseShiftInCavityWithAtom, PhaseSwitchInCavityWithAtom, TrappedIons, SinglePhotonWithScQubit,SuperpositionInScResonator, ArtificialKerrWithScQubit}, or from discrete measurements \cite{ThreePhotonSuperposition, Gkp, EmulatingCubicNonlinearity, Hybrid}. These two general approaches also differ with respect to quantum systems for which they can be applied. While the interaction with discrete variable systems is best realized by a standing wave mode in a resonator, the discrete projective measurements work better with traveling light. And here comes another distinction. The traveling modes of light are much more suitable for implementation of Gaussian operations. This is significant, because the non-Gaussian resources are useful only when the Gaussian tools are refined enough to operate without a hitch. To present a specific example, consider the issue of universal quantum information processing. In CV world this means the ability to implement a unitary operation with an arbitrary Hamiltonian \cite{QcOverCv,UniversalCvQc}. For this we need to have access to the cubic operation --- a quantum operation with Hamiltonian composed of third power of quadrature operators --- \emph{as well as} the complete range of Gaussian operations.

The Gaussian states, operations, and measurements are the foundations on which the CV quantum information processing is built. Homodyne detection, squeezed states and Gaussian linear operations in the form of displacement and passive linear optics are already staples of the contemporary experimental practice. The measurement induced paradigm \cite{MeasurementInduced}, which employs the passive linear optics together with squeezed states and linear feed-forward, then in turn allowed implementation of the Gaussian nonlinear operations such as squeezing \cite{UniversalSqueezer, SinglePhotonSqueezing}, quantum non-demolition interaction \cite{QndSum, NonlocalQnd}, and others \cite{PhaseInsensitiveAmp, QuadraticPhaseGate, FourModeGaussian}.  All these past implementations have one thing in common. The nonlinearity was static. This is not too big of a problem for the contemporary proof-of-principle experiments that are built to implement a single specific task. However, in order to move towards universal and \textit{fast} information processing, we need operations with bandwidth higher than what is allowed by `manual' change of optical elements.
The most immediate examples are the proposed experimental implementation of the cubic gate \cite{Gkp, WeakCubic} and the experimental preparation of the cubic state \cite{Gkp, SqueezingCubic}, both of which can be considered to be the important first step towards universal quantum information processing.
These applications require a nonlinear feed-forward --- a squeezing operation whose strength and direction depend on measurement results.
In this Letter we present the experimental realization of such the operation for a mode of traveling light.
This ensures that the operation can be used as a part of a larger information processing network, for example as a feed-forward in the implementation of a cubic phase gate \cite{Gkp, WeakCubic}.

%%%%%%%%%%%%%%%%%%%%%%%%%%%%%%%%%%%%%%%%%%%%%%%%%%%%%%%%%%%%%%%%%%%%%%%%%%%%%%%%
\begin{figure*}
\centering
\includegraphics{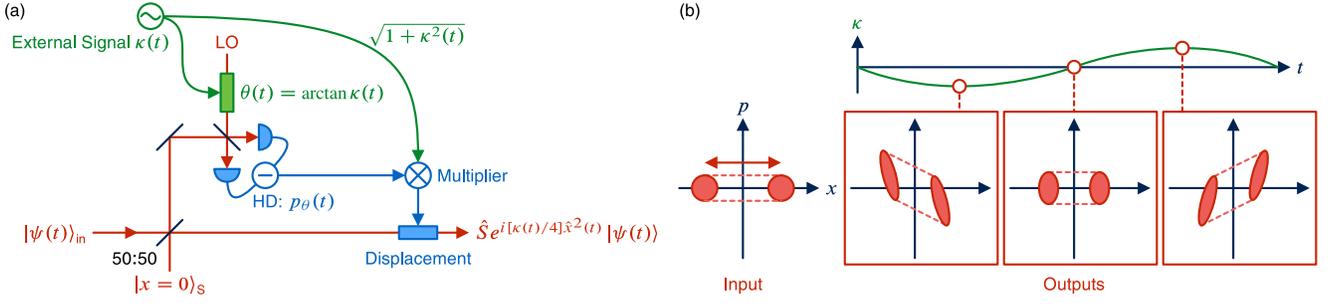}
\caption{(color online)
(a) Schematic diagram of the dynamic squeezing operation.
$\hat{S}$ denotes 3-dB squeezing of the $\hat{x}$-quadrature.
(b) Illustration of input-output relation of the experiment.
Quantum states are depicted as ellipses in phase-space representation.
}
\label{dsgschematic.eps}
\end{figure*}
%%%%%%%%%%%%%%%%%%%%%%%%%%%%%%%%%%%%%%%%%%%%%%%%%%%%%%%%%%%%%%%%%%%%%%%%%%%%%%%%

The implemented operation is a time-dependent nonlinear Gaussian operation
with an effective Hamiltonian $\hat{H}(t) = \kappa(t) \hat{x}_\text{in}^2(t)$.
This operation is in each instant applied to a different input quantum state
$\ket{\psi(t)}_\text{in}$.
Here $\kappa(t)$ is the strength of the quadratic operation, and
$\hat{x}_\text{in}(t)$ is the quadrature operator of $\ket{\psi(t)}_\text{in}$.
The operation transforms the pair of quadrature operators as
$\hat{x}(t)\to\hat{x}(t)$, $\hat{p}(t)\to\hat{p}(t)+\kappa(t)\hat{x}(t)$,
which can be decomposed into a sequence of a phase shift, a squeezing, and another phase shift (see Supplemental Material \cite{supplement}).
In this Letter we employ a streamlined experimental configuration that implements the desired transformation up to a constant local squeezing, that could be effectively compensated by the existing methods \cite{UniversalSqueezer, SinglePhotonSqueezing}.
The scheme is depicted in Figure~\ref{dsgschematic.eps}(a) and in the ideal case it works in the following way:
After we combine the input state $\ket{\psi(t)}_\text{in}$ with the $\hat{x}$-eigenstate $\ket{x=0}_\text{S}$ at a balanced beamsplitter, we measure the quadrature $\hat{p}_\theta(t) = \hat{p}_\text{HD}(t) \cos \theta(t) + \hat{x}_\text{HD}(t) \sin \theta(t)$ of one of the modes
by controlling the phase of the local oscillator (LO) of homodyne detection (HD).
Here $\theta(t)$ depends on the external driving signal as $\theta(t) = \arctan\kappa(t)$.
We then use the measured value $p_\theta(t)$ to apply $\hat{p}$-displacement to the unmeasured mode with electronic gain of $\sqrt{1 + \kappa^2(t)}$.
This transforms the quadratures of the output quantum state to
$\hat{x}(t) = \hat{x}_\text{in}(t)/\sqrt{2} $ and $\hat{p}(t) = \sqrt{2}\hat{p}_\text{in}(t) + [\kappa(t)/\sqrt{2}]\hat{x}_\text{in}(t)$.
With exception of the constant 3-dB squeezing, which can be efficiently compensated \cite{SinglePhotonSqueezing}, this is exactly the desired form.
In reality, we need to approximate the $\hat{x}$-eigenstate $\ket{x=0}_\text{S}$ with a squeezed vacuum state that can be for our purposes completely characterized by its $\hat{x}$-quadrature variance $V_x(t)$. Eventually we can derive the actual input-output relations to be
\begin{subequations}
\begin{align}
\hat{x}(t) &= \frac{1}{\sqrt{2}} \hat{x}_\text{in}(t)
              - \frac{1}{\sqrt{2}} \hat{x}_\text{S}(t), \\
\hat{p}(t) &= \sqrt{2}
              \left[\hat{p}_\text{in}(t)
                    + \frac{\kappa(t)}{2} \hat{x}_\text{in}(t) \right]
              + \frac{\kappa(t)}{\sqrt{2}} \hat{x}_\text{S}(t),
\end{align}
\label{NffIo}%
\end{subequations}
where $\hat{x}_\text{S}(t)$ denotes the $\hat{x}$-quadrature of the squeezed vacuum state and vanishes in the limit of infinite squeezing represented by $V_x(t) \rightarrow 0$.
For the sake of brevity, from now on we will be dropping the explicit notion of time-dependence of $\kappa$, $\theta$ and other operators.

The input-output relations \eqref{NffIo} can be verified by applying the operation to a set of coherent states with differing amplitudes.
In our experiment, we have chosen our input to consist of $\hat{x}$-displaced coherent states, whose amplitudes were changed in time. This allowed us to analyze the dynamic behavior with respect to both the gate parameter and the input state. In practical scenarios we can assume that the control signal is changing more slowly than the input state.
If we take one such short interval in which the control signal is constant relative to the fluctuations of the input state, the output state behaves as is depicted in Fig.~\ref{dsgschematic.eps}(b).
When $\kappa$ is around zero, the signal state is simply squeezed in the $x$-direction, and has zero mean-amplitude along the $p$-axis.
When $\kappa$ is nonzero in the time interval, the state is displaced in the $p$-direction proportionally to its initial displacement in the $x$-direction, and it is also squeezed.
The amount and the direction of the squeezing both depend on the value of $\kappa$.

%%%%%%%%%%%%%%%%%%%%%%%%%%%%%%%%%%%%%%%%%%%%%%%%%%%%%%%%%%%%%%%%%%%%%%%%%%%%%%%%
\begin{figure*}
\centering
\includegraphics{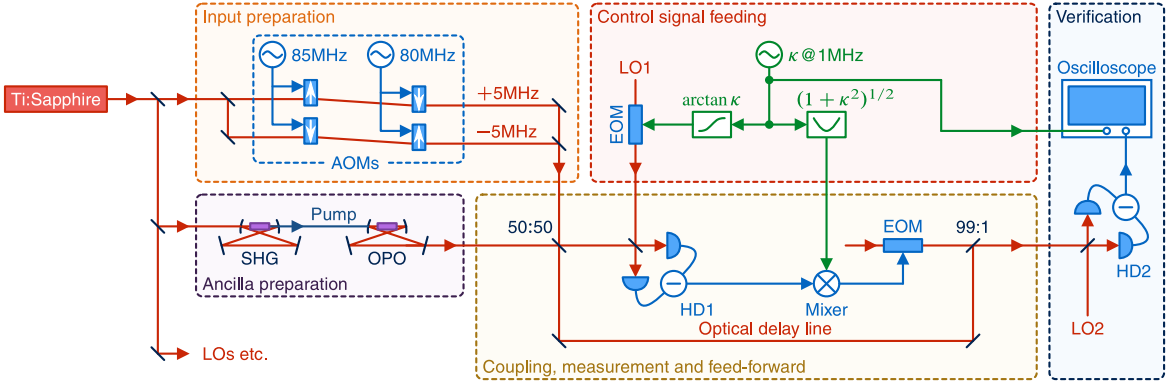}
\caption{(color online)
Experimental setup.
}
\label{setup.eps}
\end{figure*}
%%%%%%%%%%%%%%%%%%%%%%%%%%%%%%%%%%%%%%%%%%%%%%%%%%%%%%%%%%%%%%%%%%%%%%%%%%%%%%%%

The design of our experimental setup is depicted in Fig.~\ref{setup.eps}.
The light source is a continuous-wave Ti:Sapphire laser operating at 860 nm.
The input coherent state is generated at $\pm$5 MHz around the source-laser frequency with four acousto-optic modulators (AOM).
By properly locking relative phases between the frequency-shifted beams, the coherent state is displaced continuously at 5 MHz in the direction of the $x$-axis.
(This technique was previously employed in the experiment of Ref.~\cite{OpticalPhaseTracking}.)
On the other hand, the ancillary squeezed state is prepared by an optical parametric oscillator (OPO).
This OPO is a bow-tie-shaped cavity of 300 mm in length, containing a periodically-poled KTiOPO$_4$ crystal to obtain second-order nonlinearity.
The OPO is pumped by a beam with the wavelength of 430 nm and the power of 120 mW, which is generated by another bow-tie-shaped cavity (SHG) containing a KNbO$_3$ crystal.
Bandwidth of the OPO is 12.5 MHz in terms of half-width at half-maximum (HWHM) so that our setup sufficiently covers bandwidth of the input coherent state.
The typical squeezing level from DC to 10 MHz was $-$3.1 dB.
After the state preparations, we couple the input and the squeezed vacuum at a balanced beamsplitter (50:50), measuring one port of the outputs by a homodyne detector (HD1).
The measured value is used for the feed-forward system, in which the beam of the other port is suitably displaced in the $p$-direction with an electro-optic modulator (EOM), an auxiliary beam and a slightly-transmitting beamsplitter (99:1).
To match propagation times of the measured signal and the unmeasured optical beam, an optical delay line of 13 m in free space is used.
The beam-pointing of the delay line is stabilized by a piezo-actuated optical mount with a feedback system.

We have a system of feeding a control signal $\kappa$, followed by two nonlinear electronic circuits to produce $\arctan\kappa$ and $\sqrt{1+\kappa^2}$.
Here we use a sine wave with the frequency of 1 MHz as the control signal $\kappa$.
In the measurement process at HD1, the phase $\theta$ of the local oscillator (LO1) is controlled by an EOM to follow the signal $\arctan\kappa$.
The measured signal is then amplified by a factor of $\sqrt{1+\kappa^2}$.
For more details on the electronic circuits, which are key components of the dynamic gate, see the Supplemental Material \cite{supplement}.

To characterize the output states, we employ another homodyne detector (HD2).
Since the initial states are all Gaussian states and the operation is quadratic,
the output state is expected to be also Gaussian.
Therefore, to characterize the output state by homodyne detection, it is enough to see the mean values and variances of three different bases:
the $x$-axis ($\hat{x}$), the $p$-axis ($\hat{p}$), and the angle of $\pi/4$ from the $x$-axis ($\hat{x}_{\pi/4}$).
The mean values and variances of the output quadratures are obtained from the repeated measurements of 10,851 times.
The control signal $\kappa$ for each measurement is collected together.

%%%%%%%%%%%%%%%%%%%%%%%%%%%%%%%%%%%%%%%%%%%%%%%%%%%%%%%%%%%%%%%%%%%%%%%%%%%%%%%%
\begin{figure}
\centering
\includegraphics{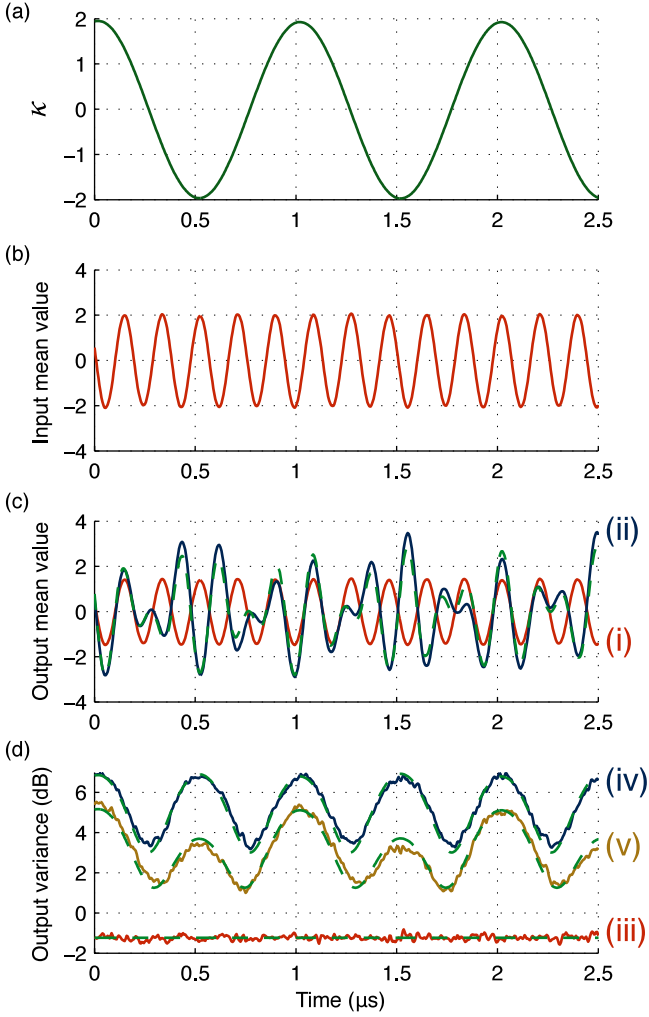}
\caption{(color online)
Experimental mean values and variances compared with theoretical predictions ($\hbar = 1$).
(a) Supplied control signal.
(b) Mean $\hat{x}$-quadrature values $\moment{\hat{x}_\text{in}}$ of the input coherent states.
Mean $\hat{p}$-quadrature values are omitted because they are always zero.
(c) Mean quadrature values of the output.
(i) $\moment{\hat{x}}$, (ii) $\moment{\hat{p}}$.
(d) Variances of the output quadratures relative to that of the shot noise.
(iii) $\moment{\varDelta \hat{x}^2}$,
(iv) $\moment{\varDelta \hat{p}^2}$,
(v) $\moment{\varDelta \hat{x}_{\pi/4}^2}$.
Solid curves are experimental results while dashed curves are theoretical predictions.
}
\label{normresult.eps}
\end{figure}
%%%%%%%%%%%%%%%%%%%%%%%%%%%%%%%%%%%%%%%%%%%%%%%%%%%%%%%%%%%%%%%%%%%%%%%%%%%%%%%%

Figure~\ref{normresult.eps} shows the experimental mean values and variances (normalized as $\hbar = 1$).
All the results are plotted in the same time domain.
Figure~\ref{normresult.eps}(a) represents the supplied control signal $\kappa$
at 1 MHz.
Figure~\ref{normresult.eps}(b) shows the mean $\hat{x}$-quadrature values $\moment{\hat{x}_\text{in}}$ of the input states, continuously fluctuating at 5 MHz.
The values in Fig.~\ref{normresult.eps}(b) were measured after the balanced beamsplitter, whose attenuation is compensated numerically by multiplying by $\sqrt{2}$.
The mean values of the $p$-quadrature are confirmed to be zero before the measurement.
Figure~\ref{normresult.eps}(c) shows the mean values $\moment{\hat{x}}$ and $\moment{\hat{p}}$ of the output states.
From Eq.~\eqref{NffIo}, $\moment{\hat{x}}$ should be independent of $\kappa$, while $\moment{\hat{p}}$ proportional to $\kappa\moment{\hat{x}_\text{in}}$.
As expected, the oscillation of $\moment{\hat{p}}$ behaves in-phase or out-of-phase with the oscillation of $\moment{\hat{x}}$ in accordance with whether $\kappa$ is positive or negative, and vanishes when $\kappa$ is zero.
Similarly, as seen in Fig.~\ref{normresult.eps}(d), the variances of the $\hat{x}$-quadrature are constantly squeezed by $-$1.3 dB, while the variances of the $\hat{p}$-quadrature oscillate at twice the frequency of the control signal $\kappa$ in accordance with the relation $\moment{\varDelta \hat{p}^2} = 2 \moment{\varDelta \hat{p}_\text{in}^2} + (\kappa^2/2)\moment{\varDelta \hat{x}_\text{in}^2}$.
Including the variances of $\hat{x}_{\pi/4}$-quadrature, those characteristics well agree with theoretical predictions plotted with dashed curves.

%%%%%%%%%%%%%%%%%%%%%%%%%%%%%%%%%%%%%%%%%%%%%%%%%%%%%%%%%%%%%%%%%%%%%%%%%%%%%%%%
\begin{figure}
\centering
\includegraphics{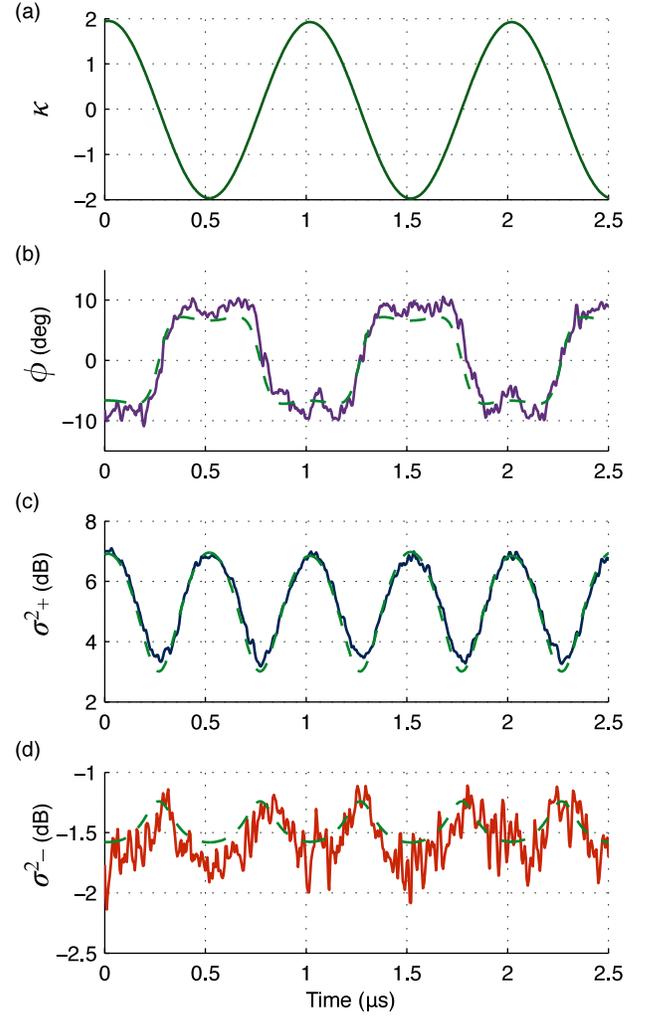}
\caption{(color online)
Analysis of the output states in terms of their diagonalized variance matrices.
(a) Supplied control signal.
(b) Squeezing angles of the output states.
(c,d) Maximally-antisqueezed variances and maximally-squeezed variances relative to the shot noise variance, respectively.
Solid curves are experimental results calculated from those in Fig.~\ref{normresult.eps}, while dashed curves are theoretical predictions.
}
\label{diagresult.eps}
\end{figure}
%%%%%%%%%%%%%%%%%%%%%%%%%%%%%%%%%%%%%%%%%%%%%%%%%%%%%%%%%%%%%%%%%%%%%%%%%%%%%%%%

We have also analyzed the output states in terms of their squeezing, both in the magnitude and the direction.
For each individual time window we have reconstructed the variance matrices of the output states as
\begin{subequations}
\begin{align}
V &= \begin{pmatrix} \sigma_x^2  & \sigma_{xp} \\
                     \sigma_{xp} & \sigma_p^2  \end{pmatrix}, \label{varmatrix} \\
\sigma_{xp}
  &= \frac{1}{2} \moment{\hat{x}\hat{p}+\hat{p}\hat{x}}
   = \sigma_{\pi/4}^2 - \frac{1}{2} (\sigma_x^2+\sigma_p^2),
\end{align}
\end{subequations}
where $\sigma_x^2 = \moment{\varDelta \hat{x}^2}$,
$\sigma_p^2 = \moment{\varDelta \hat{p}^2}$ and
$\sigma_{\pi / 4}^2 = \moment{\varDelta \hat{x}_{\pi / 4}^2}$ are the variances directly obtained from the measured data.
The variances of the squeezed and the antisqueezed quadratures, which are denoted by $\sigma_-^2$ and $\sigma_+^2$, respectively, are then found as the eigenvalues of the variance matrix (\ref{varmatrix}) :
\begin{subequations}
\begin{align}
\sigma_+^2
  &= \sigma_x^2\sin^2\phi + \sigma_p^2\cos^2\phi
     + 2\sigma_{xp}\sin\phi\cos\phi, \\
\sigma_-^2
  &= \sigma_x^2\cos^2\phi + \sigma_p^2\sin^2\phi
     - 2\sigma_{xp}\sin\phi\cos\phi, \\
\phi
  &= \frac{1}{2}
     \arctan\left(\frac{-2\sigma_{xp}}{\sigma_x^2-\sigma_p^2}\right).
\end{align}
\label{maxminvar}%
\end{subequations}
Here the parameter $\phi$ determines the direction of the squeezing, with $\phi=0$ describing the situation in which $\hat{x}$-quadrature is squeezed.
We have compared the values \eqref{maxminvar} obtained from the experimental data with the theoretical predictions and the results can be seen in Fig.~\ref{diagresult.eps}.
Figure~\ref{diagresult.eps}(a) represents again the supplied control signal $\kappa$.
Figure~\ref{diagresult.eps}(b) shows the angles of the squeezing axes $\phi$.
The square-wave-like behavior of the resulting angles means that the output states are properly rotated in phase space.
Figures~\ref{diagresult.eps}(c,d) show the maximally-antisqueezed variances $\sigma_+^2$ and the maximally-squeezed variances $\sigma_-^2$, respectively.
The maximal antisqueezing starts from about 3 dB where the control signal vanishes, and it reaches about 7 dB with $\kappa = \pm 2$.
For the maximal squeezing, it starts from about $-$1.3 dB and reaches about $-$1.8 dB.
While these values are reduced from those of the ideal case due to the finite squeezing of the ancillary states, they still show dependency on the control signal and well agree with theoretical predictions (dashed curves).

In conclusion, we have experimentally demonstrated a squeezing operation whose squeezing level and squeezing direction
can be continuously adjusted with an operational bandwidth of 1 MHz. This dynamic squeezing gate can allow implementation of an arbitrary dynamic Gaussian gate \cite{irreducibleresource} and it significantly expands the possibilities of teleportation-based quantum operations.
On a more immediate time scale, the squeezing gate is now ready to serve as the feed-forward part of the cubic phase gate \cite{Gkp,WeakCubic}.
Since the cubic phase state has been already experimentally realized \cite{EmulatingCubicNonlinearity}, the full implementation of the cubic phase gate required for the universal CV quantum information processing can be expected soon.

\begin{acknowledgments}
This work was partly supported by PDIS, GIA, APSA commissioned by the MEXT of Japan,
FIRST initiated by CSTP of Japan, ASCR-JSPS,
and the SCOPE program of the MIC of Japan. P.M. acknowledges Czech-Japan bilateral grant of MSMT CR, Grant No. LH13248 (KONTAKT). 
R.F. acknowledges GA14-36681G of Czech Science Foundation and EU FP7 BRISQ2 project (grant No.308803).
K.M.\ and H.O.\ acknowledge financial support from ALPS.
\end{acknowledgments}

%%%%%%%%%%%%%%%%%%%%%%%%%%%%%%%%%%%%%%%%%%%%%%%%%%%%%%%%%%%%%%%%%%%%%%%%%%%%%%%%

%%%%%%%%%%%%%%%%%%%%%%%%%%%%%%%%%%%%%%%%%%%%%%%%%%%%%%%%%%%%%%%%%%%%%%%%%%%%%%%%

\makeatletter
\balancelastpage@sw{%
  \onecolumngrid
 }{}%
\newpage
\makeatother

\onecolumngrid

%%%%%%%%%%%%%%%%%%%%%%%%%%%%%%%%%%%%%%%%%%%%%%%%%%%%%%%%%%%%%%%%%%%%%%%%%%%%%%%%

\begin{center} \large \bf
    Supplemental Material: Experimental realization of a dynamic squeezing gate
\end{center}

\section{I.{\quad}Implementation of nonlinear electronic circuits}

\begin{figure*}[h]
\centering
\includegraphics{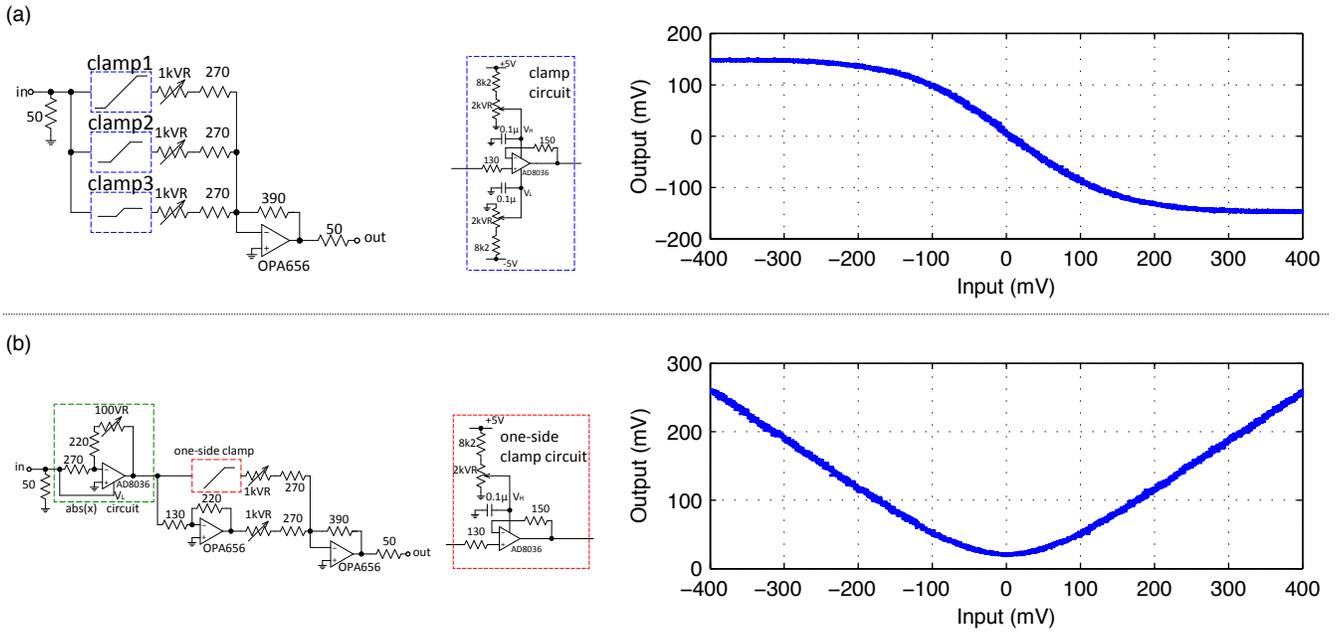}
\caption{(color online)
(Left) Schematics of nonlinear electronic circuits to process the control signal.
(Right) Experimental input-output relations of the nonlinear electronic circuits.
The input signal is a continuous triangle wave at 1 kHz.
(a) $\arctan x$ circuit.
Note that the sign of the output signal is inverted due to the inverting amplifier at the end of the circuit.
(b) $\sqrt{1 + x^2}$ circuit.
Note that the output has a fixed offset voltage, which is compensated after the circuit.
}
\label{nonlinearcircuits.eps}
\end{figure*}

As mentioned in the main text, the control signal $\kappa$ is processed to $\arctan\kappa$ and $\sqrt{1+\kappa^2}$ by nonlinear electronic circuits.
We desired to construct these nonlinear circuits with latencies as small as possible, because the rapid response is important for future application as a nonlinear feed-forward in a cubic phase gate.
For this purpose, we implemented these nonlinear processors with high-speed analog clamp circuits, as shown in Fig.~\ref{nonlinearcircuits.eps}.
In this implementation, the nonlinear functions are approximated by broken lines, and the electronic circuits work as an analog look-up table.
They operate with high precision in the range of $|\kappa| \leq 2$ at a frequency of MHz order, with the latency of less than 10 nanoseconds.

\vspace{3\baselineskip}

%%%%%%%%%%%%%%%%%%%%%%%%%%%%%%%%%%%%%%%%%%%%%%%%%%%%%%%%%%%%%%%%%%%%%%%%%%%%%%%%
%%%%%%%%%%%%%%%%%%%%%%%%%%%%%%%%%%%%%%%%%%%%%%%%%%%%%%%%%%%%%%%%%%%%%%%%%%%%%%%%

\clearpage

\section{II.{\quad}Decomposition of the quadratic operation}

As mentioned in the main text, the action of the nonlinear unitary operation with the effective Hamiltonian $\hat{H}=\kappa\hat{x}^2$ is equivalent to a sequence of a phase shift, a squeezing, and another phase shift. 
Here we mathematically show this decomposition. 
For ease in description, we introduce a new parameter $\lambda=(1/2)\arctan(\kappa/2)$, $-\pi/4<\lambda<\pi/4$. 
Then, the transformation of the quadrature operators
$(\hat{x},\hat{p})^T\to(\hat{x}^\prime,\hat{p}^\prime)^T=(\hat{x},\hat{p}+\kappa\hat{x})^T$
is calculated as follows:
\begin{align}
\begin{pmatrix}
\hat{x}^\prime \\ \hat{p}^\prime
\end{pmatrix}
&=
\begin{pmatrix}
1 & 0 \\ 2\tan2\lambda & 1
\end{pmatrix}
\begin{pmatrix}
\hat{x} \\ \hat{p}
\end{pmatrix}
\notag \\
&=
\begin{pmatrix}
\cos\lambda & -\sin\lambda \\ \sin\lambda & \cos\lambda
\end{pmatrix}
\begin{pmatrix}
\sec2\lambda & \tan2\lambda \\ \tan2\lambda & \sec2\lambda
\end{pmatrix}
\begin{pmatrix}
\cos\lambda & -\sin\lambda \\ \sin\lambda & \cos\lambda
\end{pmatrix}
\begin{pmatrix}
\hat{x} \\ \hat{p}
\end{pmatrix},
\end{align}
and therefore, this operation is a squeezing in the $\pi/4$-tilted direction sandwiched by phase shifts by $\lambda$. Note that the $\pi/4$-tilted squeezing is confirmed from the following relation, 
\begin{align}
\begin{pmatrix}
\sec2\lambda & \tan2\lambda \\ \tan2\lambda & \sec2\lambda
\end{pmatrix}
=
\begin{pmatrix}
\frac{1}{\sqrt{2}} & \frac{1}{\sqrt{2}} \\ -\frac{1}{\sqrt{2}} & \frac{1}{\sqrt{2}}
\end{pmatrix}
\begin{pmatrix}
\sec2\lambda-\tan2\lambda & 0 \\ 0 & \sec2\lambda+\tan2\lambda
\end{pmatrix}
\begin{pmatrix}
\frac{1}{\sqrt{2}} & -\frac{1}{\sqrt{2}} \\ \frac{1}{\sqrt{2}} & \frac{1}{\sqrt{2}}
\end{pmatrix}.
\end{align}
From above we know that, regarding the pure quadratic gate itself, the squeezing direction $\phi$ of the output state for a coherent-state input is $-(\pi/4)+\lambda$ when $\kappa>0$ and $(\pi/4)-\lvert\lambda\rvert$ when $\kappa<0$, having singularity at the no-squeezing $\kappa=0$. 
However, as for our experimental method, there is the additional constant squeezing as in (1), and this as well as the finite squeezing in the ancillary state made the squeezing direction continuous as shown in Fig. 4(b).

\end{document}